\documentclass[prl,aps,twocolumn,showpacs]{revtex4}
\usepackage{graphicx}
\usepackage{color}
\usepackage{epsf}
\usepackage{amsmath}
\usepackage{amssymb}
\usepackage[english]{babel}
\usepackage{bm}
\usepackage{times}
\usepackage[T1]{fontenc}
\usepackage{graphics}
\usepackage{color}
\usepackage{dcolumn}   
\usepackage{sidecap}

\newcommand {\xxi} {\ensuremath{{\bf \xi}}}

\newcommand{\bes}{ \begin{equation} \begin{split} }
\newcommand{\ees}{ \end{split} \end{equation} }

\begin{document}

\title{Inference of Time-Evolving Coupled Dynamical Systems in the Presence of Noise}
\author{Tomislav Stankovski$^1$}%
\author{Andrea Duggento$^2$}%
\author{Peter V. E. McClintock$^1$}%
\author{Aneta Stefanovska$^1$}\email{aneta@lancaster.ac.uk}%

\affiliation{$^1$ Department of Physics, Lancaster University, Lancaster, LA1 4YB, United Kingdom}

\affiliation{$^2$ Department of Biopathology and Imaging, School of Medicine \& Surgery,
Tor Vergata University, Rome, Italy}

\date{\today}

\begin{abstract}
A new method is introduced for analysis of interactions between time-dependent
coupled oscillators, based on the signals they generate. It distinguishes
unsynchronized dynamics from noise-induced phase slips, and enables the
evolution of the coupling functions and other parameters to be followed.
It is based on phase dynamics, with Bayesian inference of the time-evolving
parameters achieved by shaping the prior densities to incorporate knowledge of
previous samples. The method is tested numerically and applied to reveal and
quantify the time-varying nature of cardiorespiratory interactions.
\end{abstract}

\pacs{
02.50.Tt, 
05.45.Xt, 
05.45.Tp, 
87.10.-e, 
87.19.Hh  
}

\keywords{Synchronization, Non-autonomous systems, Dynamical
system reduction, Bayesian inference, nonlinear time-series
analysis,}

\maketitle

The common assumption that a dynamical system under study is isolated and autonomous is never rigorously true. Furthermore, it is often a poor approximation, because the inevitable external influences may be too strong to ignore. For an oscillatory system, they can e.g.\ modify its natural frequency and/or amplitude. Much effort has therefore been made to understand non-autonomous oscillators driven from equilibrium by a variety of external forcings. A more difficult problem is faced where two or more interacting oscillatory systems are subject to external deterministic influences, a scenario that often arises in practice, e.g.\ in physiology including cellular dynamics, blood circulation, and brain dynamics. In such cases, the interacting systems (e.g.\ cardio-respiratory) are influenced by other oscillatory processes as well as by noise. Similarly, interactions at the intercellular level \cite{Koseska:10} and between subcellular components \cite{MondragonPalomino:11} are crucial to multicellular organisms. Evaluation of the interactions by analysis of physiological signals (\cite{Shiogai:10} and references therein) has proved useful in relation to a diversity of different diseases.

Granger causality \cite{Daniele:08,Barnett:09} and transfer entropy \cite{Staniek:08,Hung:08} have brought insight into the functional connectivity of systems, especially in neuroscience. Based on autoregressive and information-theoretic approaches to data-driven causal inference, these methods focus on the statistical properties of the time series by measuring the extent to which the individual components exchange information. However, these methods are designed to infer effect, not mechanism. In contrast, we consider here complex interacting systems that are oscillatory and subject to noise, and extract their dynamical properties.

Several questions immediately arise in relation to the dynamics of coupled
systems. Does the external influence alter their natural frequencies or
amplitudes? Are they synchronized, or do they exhibit finite coherence? If
synchronized, is it continuous or only for some of the time? Measurements may
be relatively straightforward, using modern sensors and digital signal
acquisition equipment, but how are the resultant signals to be analysed to
reveal the characteristics of the originating systems? To date, this inverse
problem has no solution.

Earlier work on coupled oscillators emphasized the detection of synchronization \cite{Tass:98,Mormann:00,Schelter:06,Xu:06}, and quantifying the couplings and directionality of influence between the oscillators \cite{Palus:03a,Rosenblum:01,Bahraminasab:08,Jamsek:10}. The inference of an underlying phase model enabled extraction of the phase-resetting curves, interactions and structures of networks
\cite{Galan:05,Kiss:05,Miyazaki:06,Kralemann:07,Tokuda:07,Levnajic:11}. However, these techniques inferred neither the noise dynamics nor the parameters characterizing the noise. In a quite separate line of development, Bayesian inference
\cite{Smelyanskiy:05a,Friston:02,Sudderth:10,Luchinsky:08, Duggento:08,Penny:09} has opened the door to the analysis of noisy time-evolving phase dynamics.

In this Letter we introduce a new method that
(a) encompasses time-variable dynamics, (b) detects synchronization where it
exists, and (c) determines the inter-oscillator coupling functions regardless
of whether or not they are time varying. By reconstructing the dynamics in
terms of a set of base functions, we evaluate the probability that they are
driven by a set of equations that are intrinsically synchronized,
distinguishing phase-slips of dynamical origin from those attributable to
noise.
The Bayesian probability lying at the core of the
method is itself time-dependent via the prior probability as a time-dependent
informational process. Thus relatively small windows can provide good time-resolved inference.

When two noisy, weakly-interacting,  $N$-dimensional, self-sustained
oscillators synchronize \cite{Pikovsky:01}, their motion is described by their
phase dynamics:
\begin{equation}
\dot \phi_i= \omega_i + f_i(\phi_i) + g_i(\phi_i,\phi_j) + \xi_i(t)
\label{eq:phi}
\end{equation}
leaving other coordinates expressed as functions of the phase: ${\bf r_i}
\equiv {\bf r_i} (\phi_i)$. $\xxi$ is a two-dimensional noise, usually assumed Gaussian and white, $\langle \xi_i(t) \xi_j(\tau)\rangle
= \delta(t-\tau) E_{ij}$ and which may, or may not be, spatially correlated. Noise can induce phase slips in a system that would
be synchronized in the noise-free limit, so evaluation of synchronization needs precise inference of $f_i$ and $g_i$, and of the noise matrix
$E_{ij}$. The systems' periodic nature suggests periodic base-functions,
whence the use of Fourier terms for the decomposition:
\begin{equation}
\begin{split}
f_i(\phi_i) &= \sum_{k=-\infty }^ \infty \tilde{c}_{i,2k} \sin(k\phi_i) + \tilde{c}_{i,2k+1}\cos(k\phi_{i}) \\
g_i(\phi_i,\phi_j) &=   \sum_{s=-\infty }^\infty  \sum_{r=-\infty }^\infty \tilde{c}_{i;r,s}\, e^{i2\pi r \phi_i}  e^{i2\pi s \phi_j}. 
\label{eq:fourierdecomposition}
\end{split}
\end{equation}

\noindent Assuming that the dynamics is adequately described by a finite number
$K$ of Fourier terms, we can rewrite the phase dynamics of (\ref{eq:phi}) as a
finite sum of base functions:
\begin{equation}
\begin{split}
\dot \phi_l=& \sum_{k=-K}^{K} c^{(l)}_k \,
\Phi_{l,k}(\phi_1,\phi_2)  + \xxi_l(t), \label{eq:phiF}
\end{split}
\end{equation}
where $l=1,2$, $\Phi_{1,0}=\Phi_{2,0}=1$, $c^{(l)}_0=\omega_l$, and other
$\Phi_{l,k}$ and $c^{(l)}_k$ are the $K$ most important Fourier components.

In order to reconstruct the parameters of (\ref{eq:phiF}) we exploit the
approach already presented in \cite{Luchinsky:08,Duggento:08} assuming that a
2-dimensional time-series of observational data ${\mathcal X} = \{ {\bf
\phi}_{l,n} \equiv \phi_l(t_{n}) \}$ ($t_n=nh$) is provided, and that the
unknown model parameters ${\mathcal M}=\{ c^{(l)}_k , E_{ij}\}$ are to be
inferred.

In Bayesian statistics a given \emph{prior} density $p_{\mbox{\scriptsize
prior}}(\mathcal M)$ that encloses expert knowledge of the unknown parameters (based on previous observations)
and the \emph{likelihood} function $\ell ( \mathcal X | \mathcal M )$, the
probability density to observe $\{\phi_{l,n}(t)\}$ given choice $\mathcal M$ of
the dynamical model, are used to calculate the so-called \emph{posterior}
density $p_{{\mathcal X}}({\mathcal M}|{\mathcal X})$ of the unknown parameters
${\mathcal M}$, conditioned on observations, by application of Bayes' theorem $
p_{{\mathcal X}}(\mathcal M | \mathcal X) = \ell ( \mathcal X | \mathcal M ) \,
p_{\mbox{\scriptsize prior}}(\mathcal M) / \int{\ell ( \mathcal X | \mathcal M
) \, p_{\mbox{\scriptsize prior}}(\mathcal M) d \mathcal M}$.

For independent white Gaussian noise sources, and in the mid-point
approximation where $\dot \phi_{l,n}=\frac{{\phi}_{l,{n+1}}-\phi_{l,n}}{h}$ and
${\phi}_{l,n}^{\ast} = (\phi_{l,n} + \phi_{l,n+1})/2$, the likelihood is given
by a product over $n$ of the probability of observing $\phi_{l,{n+1}}$ at each
time. The negative log-likelihood function $S=-\ln \ell({\mathcal X}|{\mathcal
M})$ is
 \begin{equation}
\begin{split}
    S &=   \frac{N}{2}\ln |{E}| + \frac{h}{2}\, \sum_{n=0}^{N-1}\Big(
     c^{(l)}_k \frac{\partial \Phi_{l,k}(\phi_{\cdot,n}) }{\partial \phi_{l}}+\\
     &+ [\dot{\phi}_{i,n} - c^{(i)}_k {\Phi}_{i,k}({\phi}_{\cdot,n}^{\ast})] {({E}^{-1})}_{ij}  [\dot{\phi}_{j,n} - c^{(j)}_k {\Phi}_{j,k}({\phi}_{\cdot,n}^{\ast})] \Big )
\nonumber
\end{split}
    \label{eq:likelihood}
\end{equation}
with implicit summation over repeated indices $k$,$l$,$i$,$j$. The
log-likelihood is a function of the Fourier coefficients of the phases. Hence
for a multivariate prior probability, the posterior probability is a
multivariate normal distribution. From \cite{Luchinsky:08,Duggento:08}, and
assuming such a distribution as a prior for parameters ${c^{(l)}_{k}}$, with
mean $\bar {c}$, and covariances ${ {\bf \Xi}^{-1}}_{\mbox{\scriptsize
prior}}$, the stationary point of S is calculated recursively from:
\begin{equation}
\begin{split}
    \label{eq:cD}
     E_{ij}  &= \frac{h}{N} \left(
 \dot{\phi}_{i,n} - c^{(i)}_k {\Phi_{i,k}}({\phi}_{\cdot,n}^{\ast})  \right)
\left(
 \dot{\phi}_{j,n} - c^{(j)}_k {\Phi_{j,k}}({\phi}_{\cdot,n}^{\ast}) \right) , \\
    {r}^{(l)}_{w}  & = {({\Xi}^{-1}_\text{prior})}^{(i,l)}_{kw} \,  {c}^{(l)}_{w} +
      h \, {\Phi_{i,k}}({\phi}_{\cdot,n}^{\ast}) \,
{(E^{-1})}_{ij} \, \dot{\phi}_{j,n} +\\
 &- \frac{h}{2} \frac{\partial \Phi_{l,k}(\phi_{\cdot,n}) }{\partial \phi_{l}}, \\
\Xi^{(i,j)}_{kw}  &= {\Xi_{\text{prior}}}^{(i,j)}_{kw}   + h \,
{\Phi_{i,k}}({\phi}_{\cdot,n}^{\ast}) \, {(E^{-1})}_{ij} \,
{\Phi_{j,w}}({\phi}_{\cdot,n}^{\ast}), 
\end{split}
\end{equation}
with implicit summation over $n = 1,\ldots,N$ and over repeated indices
$k$,$l$,$i$,$j$,$w$. The mean parameter vector of the posterior is then $
c^{(i)}_k = {({\Xi}^{-1})}^{(i,l)}_{kw} \, {r}^{(l)}_{w}$. We note that a
noninformative ``flat'' prior can be used as the initial limit of an infinitely
large normal distribution, by setting ${{\bf\Xi}}_{\text{prior}}=0$ and $\bar
c_{\mbox{\scriptsize prior}}=0$. The multivariate probability ${\mathcal
N}_{\mathcal X}(c|,\bar{c},\Xi)$ for the given time series ${\mathcal X}$
explicitly defines the probability density of each parameter set of
the dynamical system.

When the sequential data come from a stream of measurements providing multiple
blocks of information, one applies (\ref{eq:cD}) to each block. If the system
is known to be non-time-varying, then the posterior density of each block is
taken as the prior of the next one. Thus, the uncertainties in the parameters
steadily decrease with time as more data are included.

If the system has time dependence, however, the method of propagating knowledge
about the state of parameters obviously has to be refined. Our framework
prescribes the prior to be multinormal, so we synthesize our knowledge into a
squared symmetric positive definite matrix. We assume that the probability of
each parameter diffuses normally with a known diffusion matrix
$\Sigma_{\text{diff}}$. Thus, the probability density of the parameters is the
convolution of two normal multivariate distributions, $\Sigma_{\text{post}}$
and $\Sigma_{\text{diff}}$: $\Sigma_{\text{prior}}^{n+1} =
\Sigma_{\text{post}}^n + \Sigma_{\text{diff}}^n$.

The particular form of $\Sigma_{\text{diff}}$ describes which part of the
dynamical fields defining the oscillators has changed, and the size of the
change. In general $(\Sigma_{\text{diff}})_{i,j} = \rho_{ij}\sigma_i \sigma_j$,
where $\sigma_i$ is the standard deviation (SD) of the diffusion of $c_i$ in
the time window $t_w$, and $\rho_{ij}$ is the correlation between the change in
the parameters $c_i$ and $c_j$. We will consider a particular example of
$\Sigma_{\text{diff}}$: we assume there is no change of correlation between
parameters ($\rho_{ij}=\delta_{ij}$) and that each SD $\sigma_i$ is a known
fraction of the relevant parameter, $\sigma_i = p_w c_i$, where $p_w$ indicates
that the parameter $p$ refers to a window of length $t_w$.

The probability of synchronized dynamics is estimated by sampling the posterior
and evaluating its overlap with the Arnold tongue border: $p_{\text{sync}}
\equiv \int s(c)\, {\mathcal N}_{{\mathcal X}}(c|\bar c,\Xi) \, \text{d} c ,$
where $s(c)=\{1,0\}$ defines whether the parameter set $c$ is inside or outside
the synchronization region. For motion on the torus $\mathbb T^2$ defined by
the toroidal coordinate $\zeta(\phi_1(t),\phi_2(t))$, and the polar coordinate
$\psi(t)$, we consider a Poincar\'e section defined by $\zeta=0$ and assume
that $d\zeta(t)/dt |_{\zeta=0} > 0$ for any $\psi$. Thus the direction of
motion along the toroidal coordinate is the same for every point of the
section, which we would like to follow in order to check whether there is a
periodic orbit. If so, and if its winding number is zero, then the system is
synchronized; and there must be at least one other periodic orbit with one of
them being stable and the other unstable.

Solution of the dynamical system over the torus yields a map $M$: $[0,2\pi]\to
[0,2\pi]$ that defines, for each $\psi_n$ on the Poincar\'e section, the next
phase $\psi_{n+1}$ after one period of the toroidal coordinate:
$\psi_{n+1}=M(\psi_n)$. The map $M$ is continuous, periodic, and has two fixed
points (one stable and one unstable) if and only if there are two periodic
orbits for the dynamical system, i.e.\ synchronization is verified if $\psi_e$
exists such that $\psi_e=M(\psi_e)$ and $ | dM(\psi_e)/d \psi | < 1$.
To calculate $s(c)$ for any of the sampled parameter sets, we: (i) fix an
arbitrary $\zeta$ and, for any $\psi_i$, integrate (\ref{eq:phiF}) numerically
for one cycle of the toroidal coordinate, obtaining the mapped point
$M(\psi_i)$; (ii) by finite difference evaluation of $dM/d\psi$ employ a
modified version of Newton's root-finding method to find the occurrence (if
any) of $\psi$ such that $M(\psi)=\psi$. If there is a root, $s(c)=1$ is
returned, otherwise $s(c)=0$ is returned.

\begin{figure}[b]
\includegraphics[width=0.98\linewidth,angle=0]{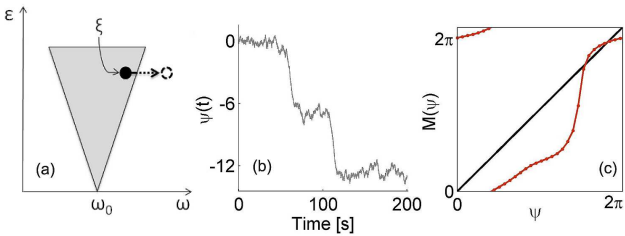}
\caption{\label{fig:phases} Synchronization discrimination for the coupled
phase oscillators (\ref{eq:phi}) with: $\omega_1=1.2$, $\omega_2=0.8$,
$\varepsilon_1=0.1$, $\varepsilon_2=0.35$, $f_i(\phi_i)=0$,
$g_1=\sin(\phi_2-\phi_1)$, $g_2=\sin(\phi_1-\phi_2)$, and noise strengths
$E_{11}=E_{22}=2$. (a) Schematic Arnold tongue to illustrate synchronization
\cite{Arnold_Tongue}. (b) Phase difference, exhibiting two phase slips. (c) Map
of $M(\psi_e)$ for (b) demonstrating that a root of $M(\psi)=\psi$ exists,
i.e.\ that the state is in fact synchronized.}
\end{figure}

From the inferred parameters of the base functions $f_i(\phi_i)$,
$g_i(\phi_i,\phi_j)$, we can reconstruct the specific functional form of the coupling functions $q_i(\phi_i,\phi_j)$. The novel
advantage of this framework is that it allows reconstruction of the
time-variability and evolution of such coupling functions. Simple normalization
of the inferred coupling parameters yields the inter-oscillator coupling
strengths, and thence the directionality index
\cite{Palus:03a,Rosenblum:01,Bahraminasab:08,Jamsek:10}. If $D\in(0,1]$ the
first oscillator drives the second ($1 \rightarrow 2$), or if $D\in[-1,0)$ otherwise. Note that, although our discussion relates to two oscillators, Eqs.\ (\ref{eq:phi})--(\ref{eq:cD}) are also applicable to a network of oscillators. For expanded discussion, technical description and software codes see \cite{bayesna_long}.

As a demonstration of how the synchronization detection works, we simulated
numerically a pair of coupled, noisy, phase oscillators (\ref{eq:phi}).
Bayesian inference followed by examination of the constructed map $M(\psi_e)$
showed that our approach successfully distinguishes synchronized ($s(c)=1$)
from unsynchronized dynamics ($s(c)=0$), i.e.\ whether the root
$M(\psi_e)=\psi_e$ exists or not. To demonstrate the novelty of our method we
consider the characteristic case illustrated in Fig.\ \ref{fig:phases}. The
parameters were such that the oscillators were only just inside the Arnold
tongue so that, for moderate noise, phase slips occurred, as shown
schematically in Fig.\ \ref{fig:phases}(a). The application of earlier methods
based on the statistics of the phase difference
\cite{Tass:98,Mormann:00,Schelter:06} suggests that the oscillators are {\it
not} synchronized. In contrast, our new technique shows that the oscillators
are {\it intrinsically} synchronized as shown in Fig.\ \ref{fig:phases}(c): the
phase slips are attributable purely to noise (the intensity of which is
inferred in matrix $E_{i,j}$), and not to deterministic interactions between
the oscillators.

\begin{figure}[t]
\includegraphics[width=0.96\linewidth,angle=0]{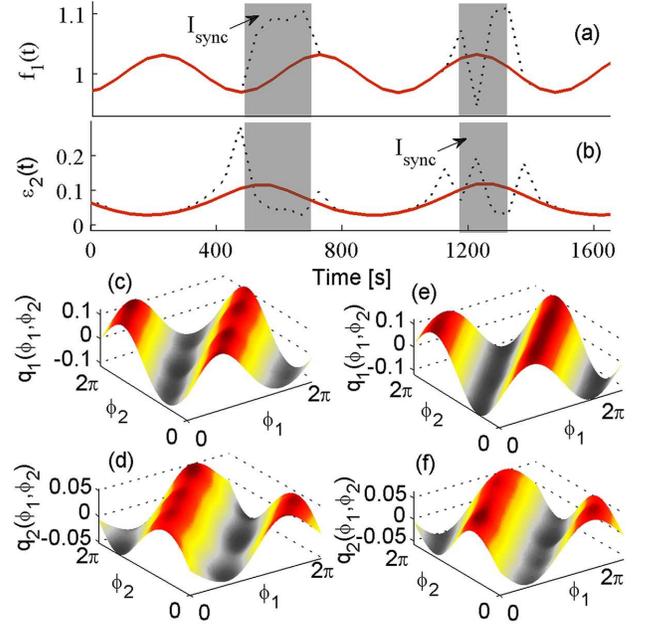}
\caption{\label{fig:num} Extraction of time-varying parameters, synchronization and coupling functions from numerical data created by (\ref{equ:num_model}). The plots show the results inferred for the numerical values of constants listed in the text. The frequency $f_1(t)$ and coupling $\varepsilon_2(t)$ are independently varied: (a) $\omega_1(t)=\omega_1+\tilde A_1\sin({\tilde \omega_1}t)$; (b) $\varepsilon_2(t)=\varepsilon_2+\tilde A_2\sin({\tilde \omega_2}t)$. The dotted and full lines plot the parameters when the two oscillators are synchronized for part of the time ($\varepsilon_1=0.3$), and not synchronized at all ($\varepsilon_1=0.1$), respectively. The regions of synchronization, found by calculation of the synchronization
index, are indicated by the gray shaded regions. (c)--(f) show the coupling functions $q_1(\phi_1,\phi_2)$ and $q_2(\phi_1,\phi_2)$ for time windows centered at different times: (c) and (d) at $t=350 s$; (e) and (f) at $t=1000 s$. The window length $t_w=50 s$, and $\varepsilon_1=0.1$ in both cases. Note the similarity in forms of (c) and (e), and of (d) and (f). The other parameters were: $\varepsilon_2=0.1$, $\omega_1=2\pi1$, $\omega_2=2\pi1.14$, $\tilde A_1=0.2$, $\tilde A_2=0.13$, $\tilde \omega_1=2\pi0.002$, $\tilde \omega_2=2\pi0.0014$ and noise $E_{11}=E_{22}=0.1$. The phases were estimated as $\phi_i = \arctan(y_i/x_i)$.}
\end{figure}

To see how the new method can also follow time-variations of the parameters,
coupling functions and synchronization, we take as an example two coupled noisy
Poincar\'e oscillators:
\begin{align}
\dot x_i&= - (\sqrt{x_i^2+y_i^2}-1)x_i  -\omega_i(t) y_i  + \varepsilon_i(t) (x_j-x_i)+\xi_i(t) \nonumber \\
\dot y_i&= - (\sqrt{x_i^2+y_i^2}-1)y_i  +\omega_i(t) x_i  + \varepsilon_i(t) (y_j-y_i)+\xi_i(t) \nonumber \\
&\text{with}\,i=1,2; \, j=1,2; \, i\neq j \, .
\label{equ:num_model}
\end{align}
We consider bidirectional coupling (1$\leftrightarrow$2), where
the natural frequency of the first oscillator, and its coupling
strength to the second one, vary periodically. For
$\varepsilon_1=0.1$ there is no synchronization: the time-varying
parameters ($f_1(t)$ and $\varepsilon_2(t)$) are accurately traced
(full red lines of Fig.\ \ref{fig:num}(a) and (b)). For a coupling
of $\varepsilon_1=0.3$ the two oscillators will be synchronized
for part of the time, resulting in intermittent synchronization.
The time-variability of the parameters in the non-synchronized
intervals is again determined correctly whereas, within the
synchronized intervals, the inferred parameters (dashed lines in
(a), (b)) diverge from their true values (full red curves). Within
these synchronized intervals, all of the base functions are highly
correlated, with values lying within the Arnold tongue. The latter
was detected as the range for which $s(c)=1$, grey-shaded in
Figs.\ \ref{fig:num}(a) and (b).

\begin{figure}[t]
\includegraphics[width=1\linewidth,angle=0]{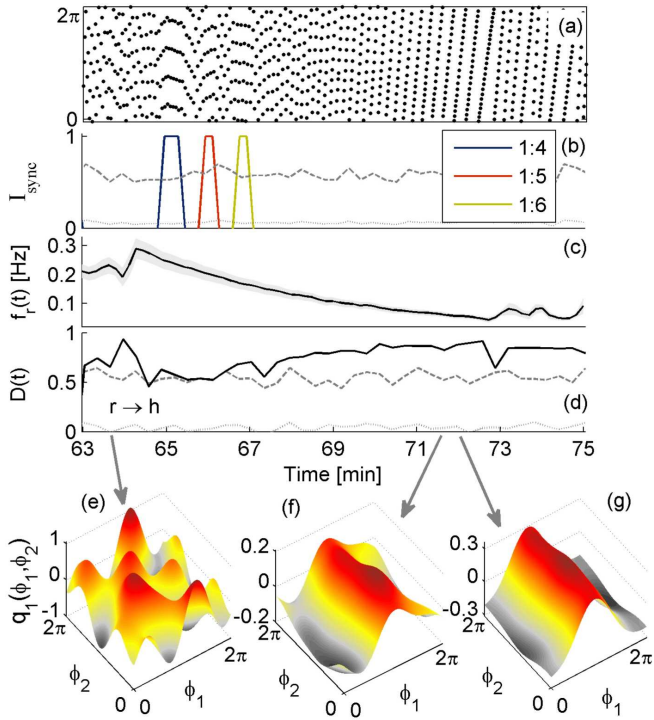}
\caption{\label{fig:CVS} Synchronization, directionality and coupling functions
in the cardio-respiratory interaction. (a) Standard 1:$N$ synchrogram.  (b)
Synchronization index for ratios 1:4, 1:5 and 1:6, as indicated. The light-gray dotted
line represents the mean, and the dark-gray dashed line the mean +2\,SD, of synchronization indices calculated from 100 surrogate \cite{Schreiber:96b} realizations. (c) The time-varying respiration
frequency (note the downward ramp due to pacing). The gray areas in (c)
represent $\pm$2\,SD from the mean value. (d) Directionality index (full curve); the light-gray dotted line represents the mean directionality index calculated from 100 surrogate realizations, and the dark-gray dashed line represents the mean +2\,SD. (e)-(g) coupling functions $q_1(\phi_1,\phi_2)$ calculated at different times, as indicated by the grey arrows.}
\end{figure}

The reconstructed sine-like functions $q_1(\phi_1,\phi_2)$ and
$q_2(\phi_1,\phi_2)$ are shown in Figs.\ \ref{fig:num}(c) and (d) for the first
and second oscillators, respectively. They describe the functional form of the
interactions between the two Poincar\'e systems (\ref{equ:num_model}).
The results suggest that the form of the coupling
functions does not evolve with time: $q_1$ and $q_2$, evaluated for later time
segments, are presented in Figs.\ \ref{fig:num} (e) and (f) respectively. By
comparison of Figs.\ \ref{fig:num}(c) and (e), or of Figs.\ \ref{fig:num}(d) and
(f), we see that the coupling functions did not change qualitatively, even
though there were time-varying parameters and weak effects from the noise.

It is well known that modulation and time-variations tend to affect
synchronization between biological oscillators
\cite{Bracic:00c,Shiogai:10,Lewis:00}. Hence the need for a technique able, not
only to identify time-varying dynamics, but also to evaluate measures of
interaction, e.g.\ synchronization, directionality and coupling
functions. To demonstrate the method on real
biological data, we analyzed cardio-respiratory measurements from resting human
subjects whose paced respiration was ramped down with decreasing frequency. The
instantaneous cardiac phase was estimated by wavelet synchrosqueezed
decomposition \cite{Daubechies:11} of the ECG signal. Similarly, the respiratory
protophase was extracted from the CO$_2$ concentration signal, followed by
transformation \cite{Kralemann:07} to the phase. The results are shown in
Fig.\ \ref{fig:CVS}. First, just for comparison, the corresponding synchrogram
\cite{Pikovsky:01} of the same data is presented in (a). The time-variation of
the respiration frequency is clearly evident in (c). By normalizing the
inferred coupling parameters, we determined the net directionality of the
interactions. Fig.\ \ref{fig:CVS}(d) suggests that the degree of directionality
is time-varying; the analyses confirm that respiration-to-heart is dominant
\cite{Shiogai:10,Palus:03a,Rosenblum:01,Bahraminasab:08,Jamsek:10}, even for
non-paced respiration (not shown). The set of inferred parameters and how they
are correlated can be used to determine whether cardiorespiratory
synchronization exists and, if so, in what ratio. Fig.\ \ref{fig:CVS}(b) shows
transitions from the non-synchronized to the synchronized state, in ratios 1:4
to 1:5 to 1:6, as the ramp progressed. The cardio-respiratory coupling
function, evaluated for three different time windows is presented in Fig.\
\ref{fig:CVS}(e)-(g). Note that the interactions are now described by complex
functions whose form changes qualitatively over time -- cf.\ Fig.\
\ref{fig:CVS}(e) with (f) and (g). This implies that, in contrast to many
systems with time-invariant coupling functions (e.g. Fig.\
\ref{fig:num}(c)--(f) or \cite{Daido:96b,Daido:96a,Crawford:95}), the
functional relations for the interactions of an open (biological) system can
themselves be \emph{time-varying processes}. By analyzing consecutive time
windows, we can even follow the time-evolution of the coupling functions --
cf.\ the similarities i.e.\ evolution of Figs.\ \ref{fig:CVS}(f) and (g).
It is important to note that the variability in form of the coupling function can cause synchronization transitions. This variability is not caused by the time-varying respiration frequency (which is decomposed separately). We also observed time-evolution of the coupling functions for spontaneous (non-paced) breathing.

In summary, our new method for inference of phase dynamics enables the
evolution of a system to be tracked continuously. Unlike earlier methods that
only detect the occurrence of transitions to/from synchronization, it reveals
details of the phase dynamics, describing the inherent nature of the
transitions and simultaneously deducing the characteristics of the noise that
stimulated them. We have identified the time-varying nature of the functions
that characterize interactions between open oscillatory systems. The
cardio-respiratory analysis demonstrated that not only the parameters, but also
the functional relationships, can be time-varying, and the new technique
follows their evolution effectively. This novel facility immediately invites
many new questions, e.g.\ the functional forms between which the couplings
vary, their frequencies of variation, how their variation affects
synchronization transitions, and whether there is periodicity or a causal
relationship waiting to be identified and understood. Thus a whole new area of
investigation has become accessible.

Our grateful thanks are due to Dwain Eckberg for providing the data for Fig.\
\ref{fig:CVS}, and to M.\ Arrayas, M.I.\ Dykman, M. Horvat, D.\ Iatsenko, P.E.\
Kloeden, D.G.\ Luchinsky, R.\ Mannella, S.\ Petkoski, and M.G.\ Rosenblum, for valuable
discussions. This work was supported by the Engineering and Physical Sciences
Research Council (UK) [grant number EP/100999X1].


\begin{thebibliography}{37}
\expandafter\ifx\csname natexlab\endcsname\relax\def\natexlab#1{#1}\fi
\expandafter\ifx\csname bibnamefont\endcsname\relax
  \def\bibnamefont#1{#1}\fi
\expandafter\ifx\csname bibfnamefont\endcsname\relax
  \def\bibfnamefont#1{#1}\fi
\expandafter\ifx\csname citenamefont\endcsname\relax
  \def\citenamefont#1{#1}\fi
\expandafter\ifx\csname url\endcsname\relax
  \def\url#1{\texttt{#1}}\fi
\expandafter\ifx\csname urlprefix\endcsname\relax\def\urlprefix{URL }\fi
\providecommand{\bibinfo}[2]{#2}
\providecommand{\eprint}[2][]{\url{#2}}

\bibitem[{\citenamefont{Koseska et~al.}(2010)\citenamefont{Koseska, Ullner,
  Volkov, Kurths, and Garcia-Ojalvo}}]{Koseska:10}
\bibinfo{author}{\bibfnamefont{A.}~\bibnamefont{Koseska}},
  \bibinfo{author}{\bibfnamefont{E.}~\bibnamefont{Ullner}},
  \bibinfo{author}{\bibfnamefont{E.}~\bibnamefont{Volkov}},
  \bibinfo{author}{\bibfnamefont{J.}~\bibnamefont{Kurths}}, \bibnamefont{and}
  \bibinfo{author}{\bibfnamefont{J.}~\bibnamefont{Garcia-Ojalvo}},
  \bibinfo{journal}{J. Theor. Biol.} \textbf{\bibinfo{volume}{263}},
  \bibinfo{pages}{189} (\bibinfo{year}{2010}).

\bibitem[{\citenamefont{{Mondragon-Palomino}
  et~al.}(2011)\citenamefont{{Mondragon-Palomino}, Danino, Selimkhanov,
  Tsimring, and Hasty}}]{MondragonPalomino:11}
\bibinfo{author}{\bibfnamefont{O.}~\bibnamefont{{Mondragon-Palomino}}},
  \bibinfo{author}{\bibfnamefont{T.}~\bibnamefont{Danino}},
  \bibinfo{author}{\bibfnamefont{J.}~\bibnamefont{Selimkhanov}},
  \bibinfo{author}{\bibfnamefont{L.}~\bibnamefont{Tsimring}}, \bibnamefont{and}
  \bibinfo{author}{\bibfnamefont{J.}~\bibnamefont{Hasty}},
  \bibinfo{journal}{Science} \textbf{\bibinfo{volume}{333}},
  \bibinfo{pages}{1315} (\bibinfo{year}{2011}).

\bibitem[{\citenamefont{Shiogai et~al.}(2010)\citenamefont{Shiogai,
  Stefanovska, and McClintock}}]{Shiogai:10}
\bibinfo{author}{\bibfnamefont{Y.}~\bibnamefont{Shiogai}},
  \bibinfo{author}{\bibfnamefont{A.}~\bibnamefont{Stefanovska}},
  \bibnamefont{and} \bibinfo{author}{\bibfnamefont{P.~V.~E.}
  \bibnamefont{McClintock}}, \bibinfo{journal}{Phys.\ Rep.}
  \textbf{\bibinfo{volume}{488}}, \bibinfo{pages}{51} (\bibinfo{year}{2010}).

\bibitem[{\citenamefont{Marinazzo et~al.}(2008)\citenamefont{Marinazzo,
  Pellicoro, and Stramaglia}}]{Daniele:08}
\bibinfo{author}{\bibfnamefont{D.}~\bibnamefont{Marinazzo}},
  \bibinfo{author}{\bibfnamefont{M.}~\bibnamefont{Pellicoro}},
  \bibnamefont{and}
  \bibinfo{author}{\bibfnamefont{S.}~\bibnamefont{Stramaglia}},
  \bibinfo{journal}{Phys. Rev. Lett.} \textbf{\bibinfo{volume}{100}},
  \bibinfo{pages}{144103} (\bibinfo{year}{2008}).

\bibitem[{\citenamefont{Barnett et~al.}(2009)\citenamefont{Barnett, Barrett,
  and Seth}}]{Barnett:09}
\bibinfo{author}{\bibfnamefont{L.}~\bibnamefont{Barnett}},
  \bibinfo{author}{\bibfnamefont{A.~B.} \bibnamefont{Barrett}},
  \bibnamefont{and} \bibinfo{author}{\bibfnamefont{A.~K.} \bibnamefont{Seth}},
  \bibinfo{journal}{Phys. Rev. Lett.} \textbf{\bibinfo{volume}{103}},
  \bibinfo{pages}{238701} (\bibinfo{year}{2009}).

\bibitem[{\citenamefont{Staniek and Lehnertz}(2008)}]{Staniek:08}
\bibinfo{author}{\bibfnamefont{M.}~\bibnamefont{Staniek}} \bibnamefont{and}
  \bibinfo{author}{\bibfnamefont{K.}~\bibnamefont{Lehnertz}},
  \bibinfo{journal}{Phys. Rev. Lett.} \textbf{\bibinfo{volume}{100}},
  \bibinfo{pages}{158101} (\bibinfo{year}{2008}).

\bibitem[{\citenamefont{Hung and Hu}(2008)}]{Hung:08}
\bibinfo{author}{\bibfnamefont{Y.}~\bibnamefont{Hung}} \bibnamefont{and}
  \bibinfo{author}{\bibfnamefont{C.}~\bibnamefont{Hu}}, \bibinfo{journal}{Phys.
  Rev. Lett.} \textbf{\bibinfo{volume}{101}}, \bibinfo{pages}{244102}
  (\bibinfo{year}{2008}).

\bibitem[{\citenamefont{Tass et~al.}(1998)\citenamefont{Tass, Rosenblum, Weule,
  Kurths, Pikovsky, Volkmann, Schnitzler, and Freund}}]{Tass:98}
\bibinfo{author}{\bibfnamefont{P.}~\bibnamefont{Tass}},
  \bibinfo{author}{\bibfnamefont{M.~G.} \bibnamefont{Rosenblum}},
  \bibinfo{author}{\bibfnamefont{J.}~\bibnamefont{Weule}},
  \bibinfo{author}{\bibfnamefont{J.}~\bibnamefont{Kurths}},
  \bibinfo{author}{\bibfnamefont{A.}~\bibnamefont{Pikovsky}},
  \bibinfo{author}{\bibfnamefont{J.}~\bibnamefont{Volkmann}},
  \bibinfo{author}{\bibfnamefont{A.}~\bibnamefont{Schnitzler}},
  \bibnamefont{and} \bibinfo{author}{\bibfnamefont{H.-J.}
  \bibnamefont{Freund}}, \bibinfo{journal}{Phys. Rev. Lett.}
  \textbf{\bibinfo{volume}{81}}, \bibinfo{pages}{3291} (\bibinfo{year}{1998}).

\bibitem[{\citenamefont{Mormann et~al.}(2000)\citenamefont{Mormann, Lehnertz,
  David, and Elger}}]{Mormann:00}
\bibinfo{author}{\bibfnamefont{F.}~\bibnamefont{Mormann}},
  \bibinfo{author}{\bibfnamefont{K.}~\bibnamefont{Lehnertz}},
  \bibinfo{author}{\bibfnamefont{P.}~\bibnamefont{David}}, \bibnamefont{and}
  \bibinfo{author}{\bibfnamefont{C.~E.} \bibnamefont{Elger}},
  \bibinfo{journal}{Physica D} \textbf{\bibinfo{volume}{144}},
  \bibinfo{pages}{358} (\bibinfo{year}{2000}).

\bibitem[{\citenamefont{Schelter et~al.}(2006)\citenamefont{Schelter,
  Winterhalder, Dahlhaus, Kurths, and Timmer}}]{Schelter:06}
\bibinfo{author}{\bibfnamefont{B.}~\bibnamefont{Schelter}},
  \bibinfo{author}{\bibfnamefont{M.}~\bibnamefont{Winterhalder}},
  \bibinfo{author}{\bibfnamefont{R.}~\bibnamefont{Dahlhaus}},
  \bibinfo{author}{\bibfnamefont{J.}~\bibnamefont{Kurths}}, \bibnamefont{and}
  \bibinfo{author}{\bibfnamefont{J.}~\bibnamefont{Timmer}},
  \bibinfo{journal}{Phys. Rev. Lett.} \textbf{\bibinfo{volume}{96}},
  \bibinfo{pages}{208103} (\bibinfo{year}{2006}).

\bibitem[{\citenamefont{Xu et~al.}(2006)\citenamefont{Xu, Chen, Hu, Stanley,
  and Ivanov}}]{Xu:06}
\bibinfo{author}{\bibfnamefont{L.~M.} \bibnamefont{Xu}},
  \bibinfo{author}{\bibfnamefont{Z.}~\bibnamefont{Chen}},
  \bibinfo{author}{\bibfnamefont{K.}~\bibnamefont{Hu}},
  \bibinfo{author}{\bibfnamefont{H.~E.} \bibnamefont{Stanley}},
  \bibnamefont{and} \bibinfo{author}{\bibfnamefont{P.~C.}
  \bibnamefont{Ivanov}}, \bibinfo{journal}{Phys.\ Rev.\ E}
  \textbf{\bibinfo{volume}{73}}, \bibinfo{pages}{065201}
  (\bibinfo{year}{2006}).

\bibitem[{\citenamefont{Palu{\v{s}} and Stefanovska}(2003)}]{Palus:03a}
\bibinfo{author}{\bibfnamefont{M.}~\bibnamefont{Palu{\v{s}}}} \bibnamefont{and}
  \bibinfo{author}{\bibfnamefont{A.}~\bibnamefont{Stefanovska}},
  \bibinfo{journal}{Phys.\ Rev.\ E} \textbf{\bibinfo{volume}{67}},
  \bibinfo{pages}{055201(R)} (\bibinfo{year}{2003}).

\bibitem[{\citenamefont{Rosenblum and Pikovsky}(2001)}]{Rosenblum:01}
\bibinfo{author}{\bibfnamefont{M.~G.} \bibnamefont{Rosenblum}}
  \bibnamefont{and} \bibinfo{author}{\bibfnamefont{A.~S.}
  \bibnamefont{Pikovsky}}, \bibinfo{journal}{Phys. Rev. E.}
  \textbf{\bibinfo{volume}{64}}, \bibinfo{pages}{045202}
  (\bibinfo{year}{2001}).

\bibitem[{\citenamefont{Bahraminasab et~al.}(2008)\citenamefont{Bahraminasab,
  Ghasemi, Stefanovska, McClintock, and Kantz}}]{Bahraminasab:08}
\bibinfo{author}{\bibfnamefont{A.}~\bibnamefont{Bahraminasab}},
  \bibinfo{author}{\bibfnamefont{F.}~\bibnamefont{Ghasemi}},
  \bibinfo{author}{\bibfnamefont{A.}~\bibnamefont{Stefanovska}},
  \bibinfo{author}{\bibfnamefont{P.~V.~E.} \bibnamefont{McClintock}},
  \bibnamefont{and} \bibinfo{author}{\bibfnamefont{H.}~\bibnamefont{Kantz}},
  \bibinfo{journal}{Phys. Rev. Lett.} \textbf{\bibinfo{volume}{100}},
  \bibinfo{pages}{084101} (\bibinfo{year}{2008}).

\bibitem[{\citenamefont{Jam\v{s}ek et~al.}(2010)\citenamefont{Jam\v{s}ek,
  Palu\v{s}, and Stefanovska}}]{Jamsek:10}
\bibinfo{author}{\bibfnamefont{J.}~\bibnamefont{Jam\v{s}ek}},
  \bibinfo{author}{\bibfnamefont{M.}~\bibnamefont{Palu\v{s}}},
  \bibnamefont{and}
  \bibinfo{author}{\bibfnamefont{A.}~\bibnamefont{Stefanovska}},
  \bibinfo{journal}{Phys. Rev. E} \textbf{\bibinfo{volume}{81}},
  \bibinfo{pages}{036207} (\bibinfo{year}{2010}).

\bibitem[{\citenamefont{Gal\'an et~al.}(2005)\citenamefont{Gal\'an, Ermentrout,
  and Urban}}]{Galan:05}
\bibinfo{author}{\bibfnamefont{R.~F.} \bibnamefont{Gal\'an}},
  \bibinfo{author}{\bibfnamefont{G.~B.} \bibnamefont{Ermentrout}},
  \bibnamefont{and} \bibinfo{author}{\bibfnamefont{N.~N.} \bibnamefont{Urban}},
  \bibinfo{journal}{Phys. Rev. Lett.} \textbf{\bibinfo{volume}{94}},
  \bibinfo{pages}{158101} (\bibinfo{year}{2005}).

\bibitem[{\citenamefont{Kiss et~al.}(2005)\citenamefont{Kiss, Zhai, and
  Hudson}}]{Kiss:05}
\bibinfo{author}{\bibfnamefont{I.~Z.} \bibnamefont{Kiss}},
  \bibinfo{author}{\bibfnamefont{Y.}~\bibnamefont{Zhai}}, \bibnamefont{and}
  \bibinfo{author}{\bibfnamefont{J.~L.} \bibnamefont{Hudson}},
  \bibinfo{journal}{Phys. Rev. Lett.} \textbf{\bibinfo{volume}{94}},
  \bibinfo{pages}{248301} (\bibinfo{year}{2005}).

\bibitem[{\citenamefont{Miyazaki and Kinoshita}(2006)}]{Miyazaki:06}
\bibinfo{author}{\bibfnamefont{J.}~\bibnamefont{Miyazaki}} \bibnamefont{and}
  \bibinfo{author}{\bibfnamefont{S.}~\bibnamefont{Kinoshita}},
  \bibinfo{journal}{Phys. Rev. Lett.} \textbf{\bibinfo{volume}{96}},
  \bibinfo{pages}{194101} (\bibinfo{year}{2006}).

\bibitem[{\citenamefont{Kralemann et~al.}(2007)\citenamefont{Kralemann,
  Cimponeriu, Rosenblum, Pikovsky, and Mrowka}}]{Kralemann:07}
\bibinfo{author}{\bibfnamefont{B.}~\bibnamefont{Kralemann}},
  \bibinfo{author}{\bibfnamefont{L.}~\bibnamefont{Cimponeriu}},
  \bibinfo{author}{\bibfnamefont{M.}~\bibnamefont{Rosenblum}},
  \bibinfo{author}{\bibfnamefont{A.}~\bibnamefont{Pikovsky}}, \bibnamefont{and}
  \bibinfo{author}{\bibfnamefont{R.}~\bibnamefont{Mrowka}},
  \bibinfo{journal}{Phys. Rev. E} \textbf{\bibinfo{volume}{76}},
  \bibinfo{pages}{055201} (\bibinfo{year}{2007}).

\bibitem[{\citenamefont{Tokuda et~al.}(2007)\citenamefont{Tokuda, Jain, Kiss,
  and Hudson}}]{Tokuda:07}
\bibinfo{author}{\bibfnamefont{I.~T.} \bibnamefont{Tokuda}},
  \bibinfo{author}{\bibfnamefont{S.}~\bibnamefont{Jain}},
  \bibinfo{author}{\bibfnamefont{I.~Z.} \bibnamefont{Kiss}}, \bibnamefont{and}
  \bibinfo{author}{\bibfnamefont{J.~L.} \bibnamefont{Hudson}},
  \bibinfo{journal}{Phys. Rev. Lett.} \textbf{\bibinfo{volume}{99}},
  \bibinfo{pages}{064101} (\bibinfo{year}{2007}).

\bibitem[{\citenamefont{Levnaji\'{c} and Pikovsky}(2011)}]{Levnajic:11}
\bibinfo{author}{\bibfnamefont{Z.}~\bibnamefont{Levnaji\'{c}}}
  \bibnamefont{and} \bibinfo{author}{\bibfnamefont{A.}~\bibnamefont{Pikovsky}},
  \bibinfo{journal}{Phys. Rev. Lett.} \textbf{\bibinfo{volume}{107}},
  \bibinfo{pages}{034101} (\bibinfo{year}{2011}).

\bibitem[{\citenamefont{Smelyanskiy et~al.}(2005)\citenamefont{Smelyanskiy,
  Luchinsky, Stefanovska, and McClintock}}]{Smelyanskiy:05a}
\bibinfo{author}{\bibfnamefont{V.~N.} \bibnamefont{Smelyanskiy}},
  \bibinfo{author}{\bibfnamefont{D.~G.} \bibnamefont{Luchinsky}},
  \bibinfo{author}{\bibfnamefont{A.}~\bibnamefont{Stefanovska}},
  \bibnamefont{and} \bibinfo{author}{\bibfnamefont{P.~V.~E.}
  \bibnamefont{McClintock}}, \bibinfo{journal}{Phys.\ Rev.\ Lett.}
  \textbf{\bibinfo{volume}{94}}, \bibinfo{pages}{098101}
  (\bibinfo{year}{2005}).

\bibitem[{\citenamefont{Friston}(2002)}]{Friston:02}
\bibinfo{author}{\bibfnamefont{K.~J.} \bibnamefont{Friston}},
  \bibinfo{journal}{NeuroImage} \textbf{\bibinfo{volume}{16}},
  \bibinfo{pages}{513} (\bibinfo{year}{2002}).

\bibitem[{\citenamefont{Sudderth et~al.}(2010)\citenamefont{Sudderth, Ihler,
  Isard, Freeman, and Willsky}}]{Sudderth:10}
\bibinfo{author}{\bibfnamefont{E.~B.} \bibnamefont{Sudderth}},
  \bibinfo{author}{\bibfnamefont{A.~T.} \bibnamefont{Ihler}},
  \bibinfo{author}{\bibfnamefont{M.}~\bibnamefont{Isard}},
  \bibinfo{author}{\bibfnamefont{W.~T.} \bibnamefont{Freeman}},
  \bibnamefont{and} \bibinfo{author}{\bibfnamefont{A.~S.}
  \bibnamefont{Willsky}}, \bibinfo{journal}{Commun.\ ACM}
  \textbf{\bibinfo{volume}{53}}, \bibinfo{pages}{95} (\bibinfo{year}{2010}).

\bibitem[{\citenamefont{Luchinsky et~al.}(2008)\citenamefont{Luchinsky,
  Smelyanskiy, Duggento, and McClintock}}]{Luchinsky:08}
\bibinfo{author}{\bibfnamefont{D.~G.} \bibnamefont{Luchinsky}},
  \bibinfo{author}{\bibfnamefont{V.~N.} \bibnamefont{Smelyanskiy}},
  \bibinfo{author}{\bibfnamefont{A.}~\bibnamefont{Duggento}}, \bibnamefont{and}
  \bibinfo{author}{\bibfnamefont{P.~V.~E.} \bibnamefont{McClintock}},
  \bibinfo{journal}{Phys. Rev. E} \textbf{\bibinfo{volume}{77}},
  \bibinfo{pages}{061105} (\bibinfo{year}{2008}).

\bibitem[{\citenamefont{Duggento et~al.}(2008)\citenamefont{Duggento,
  Luchinsky, Smelyanskiy, Khovanov, and McClintock}}]{Duggento:08}
\bibinfo{author}{\bibfnamefont{A.}~\bibnamefont{Duggento}},
  \bibinfo{author}{\bibfnamefont{D.~G.} \bibnamefont{Luchinsky}},
  \bibinfo{author}{\bibfnamefont{V.~N.} \bibnamefont{Smelyanskiy}},
  \bibinfo{author}{\bibfnamefont{I.}~\bibnamefont{Khovanov}}, \bibnamefont{and}
  \bibinfo{author}{\bibfnamefont{P.~V.~E.} \bibnamefont{McClintock}},
  \bibinfo{journal}{Phys. Rev. E} \textbf{\bibinfo{volume}{77}},
  \bibinfo{pages}{061106} (\bibinfo{year}{2008}).

\bibitem[{\citenamefont{Penny et~al.}(2009)\citenamefont{Penny, Litvak,
  Fuentemilla, Duzel, and Friston}}]{Penny:09}
\bibinfo{author}{\bibfnamefont{W.~D.} \bibnamefont{Penny}},
  \bibinfo{author}{\bibfnamefont{V.}~\bibnamefont{Litvak}},
  \bibinfo{author}{\bibfnamefont{L.}~\bibnamefont{Fuentemilla}},
  \bibinfo{author}{\bibfnamefont{E.}~\bibnamefont{Duzel}}, \bibnamefont{and}
  \bibinfo{author}{\bibfnamefont{K.}~\bibnamefont{Friston}},
  \bibinfo{journal}{J. Neurosci. Methods} \textbf{\bibinfo{volume}{183}},
  \bibinfo{pages}{19} (\bibinfo{year}{2009}).

\bibitem[{\citenamefont{Pikovsky et~al.}(2001)\citenamefont{Pikovsky,
  Rosenblum, and Kurths}}]{Pikovsky:01}
\bibinfo{author}{\bibfnamefont{A.}~\bibnamefont{Pikovsky}},
  \bibinfo{author}{\bibfnamefont{M.}~\bibnamefont{Rosenblum}},
  \bibnamefont{and} \bibinfo{author}{\bibfnamefont{J.}~\bibnamefont{Kurths}},
  \emph{\bibinfo{title}{Synchronization -- A Universal Concept in Nonlinear
  Sciences}} (\bibinfo{publisher}{Cambridge University Press},
  \bibinfo{address}{Cambridge}, \bibinfo{year}{2001}).

\bibitem[{Arn()}]{Arnold_Tongue}
\emph{\bibinfo{title}{\rm {N}ote that the {A}rnold tongue in this case is
  considered to be valid for the intrinsic parameters without the effect from
  the noise. {T}he border of the {A}rnold tongue from the full dynamics
  (including the noise) might not be sharp, and can be ``blurred'' by the
  noise.}}

\bibitem[{\citenamefont{Schreiber and Schmitz}(1996)}]{Schreiber:96b}
\bibinfo{author}{\bibfnamefont{T.}~\bibnamefont{Schreiber}} \bibnamefont{and}
  \bibinfo{author}{\bibfnamefont{A.}~\bibnamefont{Schmitz}},
  \bibinfo{journal}{Phys. Rev. Lett.} \textbf{\bibinfo{volume}{77}},
  \bibinfo{pages}{635} (\bibinfo{year}{1996}).

\bibitem[{\citenamefont{Lotri{\v{c}} and Stefanovska}(2000)}]{Bracic:00c}
\bibinfo{author}{\bibfnamefont{M.~B.} \bibnamefont{Lotri{\v{c}}}}
  \bibnamefont{and}
  \bibinfo{author}{\bibfnamefont{A.}~\bibnamefont{Stefanovska}},
  \bibinfo{journal}{Physica A} \textbf{\bibinfo{volume}{283}},
  \bibinfo{pages}{451} (\bibinfo{year}{2000}).

\bibitem[{\citenamefont{Lewis et~al.}(2000)\citenamefont{Lewis, Gebber, Zhong,
  Larsen, and Barman}}]{Lewis:00}
\bibinfo{author}{\bibfnamefont{C.~D.} \bibnamefont{Lewis}},
  \bibinfo{author}{\bibfnamefont{G.~L.} \bibnamefont{Gebber}},
  \bibinfo{author}{\bibfnamefont{S.}~\bibnamefont{Zhong}},
  \bibinfo{author}{\bibfnamefont{P.~D.} \bibnamefont{Larsen}},
  \bibnamefont{and} \bibinfo{author}{\bibfnamefont{S.~M.}
  \bibnamefont{Barman}}, \bibinfo{journal}{J.\ Neurophysiol.}
  \textbf{\bibinfo{volume}{84}}, \bibinfo{pages}{1157} (\bibinfo{year}{2000}).

\bibitem[{\citenamefont{Daubechies et~al.}(2011)\citenamefont{Daubechies, Lu,
  and Wu}}]{Daubechies:11}
\bibinfo{author}{\bibfnamefont{I.}~\bibnamefont{Daubechies}},
  \bibinfo{author}{\bibfnamefont{J.}~\bibnamefont{Lu}}, \bibnamefont{and}
  \bibinfo{author}{\bibfnamefont{H.}~\bibnamefont{Wu}},
  \bibinfo{journal}{Appl.\ and Comput.\ Harmon.\ Anal.}
  \textbf{\bibinfo{volume}{30}}, \bibinfo{pages}{243} (\bibinfo{year}{2011}).

\bibitem[{\citenamefont{Daido}(1996{\natexlab{a}})}]{Daido:96b}
\bibinfo{author}{\bibfnamefont{H.}~\bibnamefont{Daido}},
  \bibinfo{journal}{Phys. Rev. Lett.} \textbf{\bibinfo{volume}{77}},
  \bibinfo{pages}{1406} (\bibinfo{year}{1996}{\natexlab{a}}).

\bibitem[{\citenamefont{Daido}(1996{\natexlab{b}})}]{Daido:96a}
\bibinfo{author}{\bibfnamefont{H.}~\bibnamefont{Daido}},
  \bibinfo{journal}{Physica D: Nonlinear Phenomena}
  \textbf{\bibinfo{volume}{91}}, \bibinfo{pages}{24 }
  (\bibinfo{year}{1996}{\natexlab{b}}).

\bibitem[{\citenamefont{Crawford}(1995)}]{Crawford:95}
\bibinfo{author}{\bibfnamefont{J.~D.} \bibnamefont{Crawford}},
  \bibinfo{journal}{Phys. Rev. Lett.} \textbf{\bibinfo{volume}{74}},
  \bibinfo{pages}{4341} (\bibinfo{year}{1995}).


\bibitem[{\citenamefont{Duggento}(2012)}]{bayesna_long}
\bibinfo{author}{\bibfnamefont{A.} \bibnamefont{Duggento}},
\bibinfo{author}{\bibfnamefont{T.} \bibnamefont{Stankovski}},
\bibinfo{author}{\bibfnamefont{P.~V.~E.} \bibnamefont{McClintock}} \bibnamefont{and}
\bibinfo{author}{\bibfnamefont{A.} \bibnamefont{Stefanovska}}
  \bibinfo{journal}{- to be published}.



\end{thebibliography}

\end{document}